\documentclass[lettersize,journal]{IEEEtran}
\usepackage{amsmath,amsfonts}
\usepackage{algorithmic}
\usepackage{array}
\usepackage[caption=false,font=normalsize,labelfont=sf,textfont=sf]{subfig}
\usepackage{textcomp}
\usepackage{xcolor}
\usepackage{color}

\usepackage{stfloats}
\usepackage{url}
\usepackage{verbatim}
\usepackage{graphicx}
\def\BibTeX{{\rm B\kern-.05em{\sc i\kern-.025em b}\kern-.08em
    T\kern-.1667em\lower.7ex\hbox{E}\kern-.125emX}}
\usepackage{balance}
\begin{document}
\title{Energy Efficient Transmitter Creation by Consuming Free Energy in Molecular Communication}
\author{Dongliang Jing, \IEEEmembership{Member, IEEE}, Linjuan Li, Zhen Cheng, Lin Lin, \IEEEmembership{Senior Member,IEEE}, and Andrew W. Eckford, \IEEEmembership{Senior Member,IEEE} 
\thanks{This work was supported by China Postdoctoral Science Foundation under Grant 2023M732877, in part by the Natural Science Basic Research Program
of Shaanxi under Program 2024JC-YBQN-0649, and in part by the National Natural Science Foundation of China under Grant 62271446.}
\thanks{Dongliang Jing is with the College of Mechanical and Electronic Engineering, Northwest A\&F University, Yangling, China, and the Key Laboratory of Agricultural Internet of Things, Ministry of Agriculture and Rural Affairs, Yangling, China (e-mail: dljing@nwafu.edu.cn).}
\thanks{Linjuan Li is with the College of Mechanical and Electronic Engineering, Northwest A\&F University, Yangling, China (e-mail: 2021012605@nwafu.edu.cn).}
\thanks{Zhen Cheng is with the School of Computer Science and Technology, Zhejiang University of Technology, Hangzhou, 310023, China (e-mail: chengzhen@zjut.edu.cn).}
\thanks{Lin Lin is with the College of Electronics and Information Engineering, Tongji University, Shanghai, China (e-mail: fxlinlin@tongji.edu.cn).}
\thanks{Andrew W. Eckford is with the Department of Electrical Engineering and Computer Science, York University, Toronto, Ontario, Canada (e-mail: aeckford@yorku.ca).}
\thanks{Part of this paper has been accepted for publication in the Asia-Pacific Molecular Communications Workshop.}}

\markboth{Accepted for publication in IEEE Transactions on Molecular, Biological, and Multi-Scale Communications - Preprint}%
{Accepted for publication in IEEE Transactions on Molecular, Biological, and Multi-Scale Communications - Preprint}

\maketitle

\begin{abstract}
Information molecules play a crucial role in molecular communication (MC), acting as carriers for information transfer. A common approach to get information molecules in MC involves harvesting them from the environment; however, the harvested molecules are often a mixture of various environmental molecules, and the initial concentration ratios in the reservoirs are identical, which hampers high-fidelity transmission techniques such as molecular shift keying (MoSK).
This paper presents a transmitter design that harvests molecules from the surrounding environment and stores them in two reservoirs. To separate the mixed molecules, energy is consumed to transfer them between reservoirs. Given limited energy resources, this work explores energy-efficient strategies to optimize transmitter performance. Through theoretical analysis and simulations, we investigate different methods for moving molecules between reservoirs. The results demonstrate that transferring higher initial concentration molecules enhances transmitter performance, while using fewer molecules per transfer further improves efficiency. These findings provide valuable insights for optimizing MC systems through energy-efficient molecule transfer techniques.
\end{abstract}


\section{Introduction}
With the rapid advancement of nanotechnology, significant breakthroughs have been achieved in the development of nanomachines, especially for applications in smart drug delivery, the early diagnosis and treatment of diseases  \cite{nakano2012molecular,farsad2016comprehensive,qiu2023review}. 
However, the constraints of nanomachines, such as limited energy and computing capacity, restrict the working capabilities of a single nanomachine. These limitations also render traditional electromagnetic communication infeasible \cite{li2019spatial}. Inspired by communication among living cells, molecular communication (MC) employs chemical signals as carriers for information transmission between transmitters and receivers and is considered one of the most promising paradigms in nanomachine technology \cite{borges2021toward,lotter2023experimental}. During the COVID-19 pandemic, MC was also employed to detect and model the spread of infections and diseases \cite{schurwanz2021infectious,chen2022detection}. 

In most previous studies, the transmitter in diffusion-based molecular communication (DBMC) was assumed to be capable of releasing the required number and types of molecules \cite{kuran2011modulation,cheng2021mobile,xiao2023really}. Concentration shift keying (CSK) and molecular shift keying (MoSK) were proposed in \cite{kuran2011modulation}. In CSK, information is encoded in the concentration levels of a single type of molecule, while MoSK uses multiple types of molecules to convey information. In \cite{xiao2023really}, the fundamental physical characteristics of commonly used information molecules, along with their associated communication systems and potential applications, were examined in detail.
In \cite{huang2019space}, Huang $et$ $al$. proposed a space shift keying-based molecular communication system in which transmitters were modeled as point sources, releasing molecules in an omnidirectional manner. In \cite{gursoy2019index}, multiple distinct point sources on the transmitter's surface acted as transmit antennas, releasing molecules according to the applied modulation scheme. Reference \cite{gao2020molecular} explored a molecular type spread MoSK scheme to enable multiple-access transmission, utilizing various types of information molecules for communication among nanomachines. In \cite{araz2023ratio}, a ratio shift keying scheme was introduced, where information is encoded in the concentration ratio of two different types of molecules. Furthermore, in \cite{civas2024frequency}, the transmitter encoded information using binary CSK by releasing a specific number of information molecules corresponding to each bit.

In DBMC, information molecules are released by a transmitter and propagate to a receiver through free diffusion. This process does not require external energy, making DBMC an energy-efficient communication scheme for nanomachines \cite{bi2021survey,yue2022cooperative}. The main energy consumption in DBMC comes from the synthesis and/or purification of information molecules and their subsequent release from the transmitter \cite{kuran2010energy}. Bio-nanomachines are capable of harvesting energy from their environment, but the energy is limited\cite{nakano2012molecular}.

To enhance energy efficiency, power adjustment techniques utilizing residual molecules in the communication channel have been proposed \cite{tepekule2015isi,jing2020power}. Additionally, an energy-efficient approach for the synthesis of information molecules has been considered, leveraging a simultaneous molecular information and energy transfer relay. In this approach, emission molecules are generated using absorbed molecules through chemical reactions \cite{deng2016enabling}.
Taking into account both energy and performance constraints, an optimized transmission scheme for DBMC has been developed, which dynamically adjusts the number of molecules released on a per-frame basis \cite{musa2022optimized}. Furthermore, research has focused on optimizing the energy efficiency of a molecular data-collection nanonetwork, comprising a mobile nanorobot and energy-constrained nanosensors, while considering constraints related to molecular concentration, data rate, and available molecular resources \cite{panahi2024energy}.

In synaptic communication, neurotransmitters released into the synaptic cleft are reabsorbed through a process called reuptake, which plays a crucial role in regulating synaptic signaling \cite{rudnick1993synapse}. Inspired by this mechanism, molecule-harvesting techniques have been proposed to increase the molecular resources available at the transmitter \cite{ahmadzadeh2022molecule,huang2023analysis,wen2024absorption,huang2024heterogeneous}. In \cite{ahmadzadeh2022molecule}, mathematical models for molecule-harvesting transmitters were introduced, where the transmitter is equipped with harvesting units on its surface to recapture molecules that come into contact with these units. Similarly, in \cite{huang2023analysis}, a transmitter model with a surface covered by heterogeneous receptors was studied, where molecules are absorbed upon hitting any receptor. Furthermore, \cite{wen2024absorption} proposed an absorption shift keying scheme to enhance system performance by employing a third, switch-controllable molecule-harvesting node to capture unused molecules. Finally, \cite{huang2024heterogeneous} considered a transmitter covered by heterogeneous receptors, which can absorb the released molecules once they hit any receptor, and analyzed the expected fraction of absorbed molecules.

In this paper, we explore a scenario where a transmitter actively collects molecules from its surroundings and then stores them in a dedicated reservoir in the transmitter itself. The types of molecules collected from the environment are not unique; instead, they consist of a mixture, as the surroundings contain multiple molecular species. To reduce the complexity of the analysis, we make a simplifying assumption that only two distinct types of molecules are harvested. These molecules are then stored in separate reservoirs within the transmitter. By focusing on just two molecular types, we can more effectively model and analyze the energy consumption involved in the process of creating the transmitter. Building on previous works \cite{eckford2018thermodynamic, jing2023thermodynamic, jing2024performance}, which focus on moving a single type of molecule between reservoirs by consuming free energy, we extend the investigation to the separation and movement of different molecule types. Given the energy constraints within nanomachines, achieving energy-efficient transmitter operation is crucial for MC. Therefore, this paper explores the performance of the transmitter under different molecule-transport strategies. 

Our primary contribution lies in the development of a highly energy-efficient mechanism for transmitter operation by optimizing various methods for transferring molecules between reservoirs. 
The main contributions of this paper are summarized as follows: 
\begin{itemize} 
\item Different molecule transport strategies between the reservoirs are studied, and the corresponding transmitter performance is analyzed. 
\item We analyze the impact of varying the number of molecules transported on the performance of the transmitter. 
\item A highly energy-efficient mechanism for transmitter creation is proposed, considering different initial states of molecule concentrations. 
\end{itemize}

The remainder of this paper is organized as follows. Section \uppercase\expandafter{\romannumeral 2} describes the considered system model. In section \uppercase\expandafter{\romannumeral 3}, the performance analysis of imperfect transmitters with different moving strategies is studied. The performance of the considered imperfect transmitter is evaluated in
\uppercase\expandafter{\romannumeral 4}, and finally, conclusions are given in Section \uppercase\expandafter{\romannumeral 5}.
\section{System Model}

In this paper, we consider a point transmitter composed of two molecular reservoirs: a low ($L$) reservoir and a high ($H$) reservoir, where information molecules are stored. Initially, information molecules are collected from the environment and uniformly distributed between the two reservoirs due to diffusion, resulting in equal concentrations of the same type of molecules in both the low and high reservoirs. The number of harvested molecules at one time is assumed to be large enough for information transmission, eliminating the need for additional harvesting during transmission. 

To make MoSK applicable, a concentration difference between the low and high reservoirs is required, allowing transmitted information to be encoded on the concentration ratio of the two molecule types. To achieve this concentration difference, molecules are moved between the reservoirs in order to purify them, which consumes free energy. However, the energy cost increases rapidly with the number of moved molecules, making it unfeasible to completely separate the two types of information molecules. Consequently, while a concentration difference between the reservoirs is achieved, the molecules remain mixed within each reservoir, resulting in the creation of an imperfect transmitter. Additionally, different moving strategies lead to varying concentration ratios of the two types of molecules between the reservoirs, which in turn affects transmitter performance.

\begin{figure}
\centering
\includegraphics[width=0.5\textwidth]{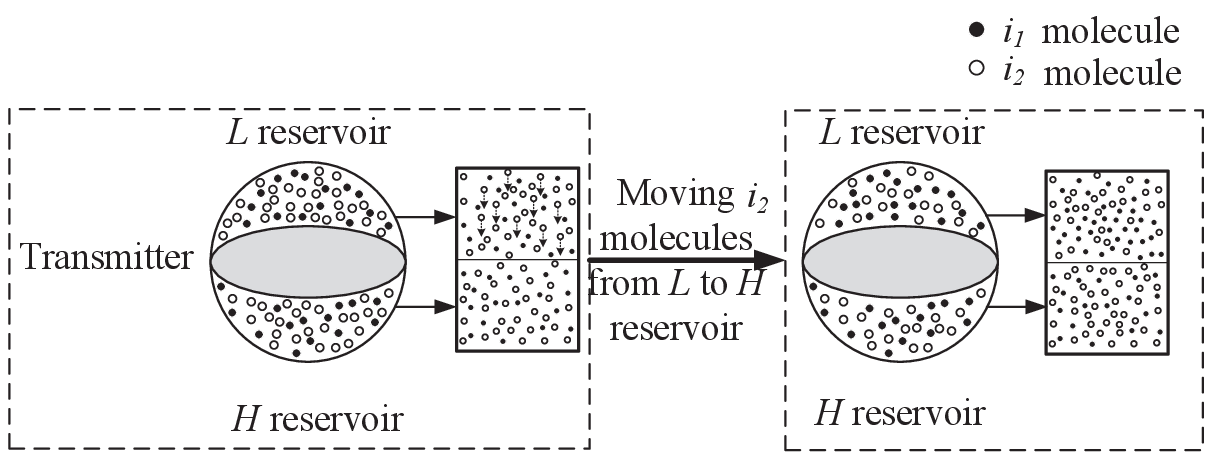}
\caption{The creation of an imperfect transmitter involves moving molecules between the reservoirs. For example, by consuming free energy, $i_2$ molecules are moved from the $L$ reservoir to the $H$ reservoir, creating a concentration difference between the transmitter reservoirs and forming an imperfect transmitter.} \label{system_model}
\end{figure}

Fig. \ref{system_model} illustrates the process of forming an imperfect transmitter. Information molecules are harvested from the environment and are assumed to start with a higher initial concentration of $i_2$ molecules in the reservoirs. By consuming free energy, $i_2$ molecules are transferred from the $L$ reservoir to the $H$ reservoir, creating a concentration difference between the two reservoirs and resulting in an imperfect transmitter. This is just an example; $i_1$ molecules can also be transferred from the $L$ reservoir to the $H$ reservoir, and both types of molecules ($i_1$ and $i_2$) can be transferred from the $H$ reservoir to the $L$ reservoir. In this MC system, information is encoded on the concentration ratio between two types of molecules. 
% To transmit bit 0, $N_m$ molecules are released from the low reservoir, while to transmit bit, $N_m$ molecules are released from the high reservoir.

At the initial state, let the concentration of $i_2$ molecules in both the $L$ and $H$ reservoirs be $c_2$, where $c_2$ is the ratio of the number of $i_2$ molecules to the total number of molecules in each reservoir. Thus, the concentration of $i_1$ molecules is $1-c_2$. Let $n_L$ and $n_H$ denote the number of molecules in the $L$ and $H$ reservoirs, respectively, so the total number of molecules in the transmitter is $n=n_L+n_H$. Assuming an equal number of molecules in both reservoirs initially, we have $n_L = n_H = \frac{1}{2}n$.

As the available free energy is limited and the initial concentrations of the two types of molecules are different, the number of molecules that can be moved varies under different moving strategies. However, the number of moved molecules directly affects the performance of the transmitter.
Assuming $m_2$ $i_2$ molecules are moved from the low reservoir to the high reservoir, and ${m_2} \ll \frac{n}{2}$, meaning the number of moved molecules is much smaller compared to the total number of molecules in the reservoirs, the number of moved $m_2$ molecules under a given energy can be expressed as \cite{ jing2023thermodynamic}
\begin{align}
\label{m_2}
{m_2} = \sqrt {\frac{{{c_2}{n}}}{{2KT}}E}.
\end{align}

Similarly, assuming $m_1$ $i_1$ molecules are moved, and ${m_1} \ll \frac{n_i}{2}$, where ${c_1} = 1 - {c_2}$, then, the number of moved $i_1$ molecules under a given energy can be expressed as
\begin{align}
\label{m_1}
{m_1} = \sqrt {\frac{{(1 - {c_2}){n}}}{{2KT}}E}.
\end{align}

From (\ref{m_2}) and (\ref{m_1}), it can be seen that the number of moved molecules $m$ is related to the input free energy $E$ and its initial concentration. Under the given energy, the number of moved molecules is determined by the initial concentration of that molecule, with a larger initial concentration resulting in a larger number of molecules that can be moved.
% Moreover, under the same conditions, $m_1$ and $m_2$ can be considered as a univariate function of the initial concentration $c_2$. It can be observed that ${m_1}(1 - c_2) = {m_2}c_2$, indicating that $m_1$ and $m_2$ are symmetric with respect to $c_2 = 0.5$.

After the imperfect transmitter is created, with molecules mixed in the reservoirs, information is encoded based on the molecular concentration ratio of different types of molecules. To transmit bit 0, $N_m$ molecules are selected from the low reservoir and released by the transmitter into the channel. Similarly, to transmit bit 1, $N_m$ molecules are selected from the high reservoir and released. Notably, the molecules selected from the reservoir are chosen randomly. In this paper, we focus on the errors introduced by the transmitter and primarily analyze its performance under different moving strategies with limited free energy.
\section{Performance analysis of imperfect transmitters with different moving strategies}
In the considered imperfect transmitter, performance is primarily determined by the concentration ratios of various types of molecules. However, different molecular transport strategies consume varying amounts of free energy, leading to distinct concentration ratios and, consequently, fluctuations in transmitter performance. Therefore, studying the relationship between transmitter performance and the molecular transport strategies employed is essential for optimizing system efficiency.

Given that the initial concentrations of $i_1$ and $i_2$ molecules are different—a reasonable assumption since information molecules stored in reservoirs can be sourced from the environment, where the concentrations of different molecule types often vary. Meanwhile, we assume that the initial fraction of $i_2$ molecules is larger than that of $i_1$ molecules.

In the initial state, the concentrations of $i_2$ molecules in both $L$ and $H$ reservoirs are $c_2$, while the concentration of $i_1$ molecules is $1-c_2$. The initial concentrations ratio  of $i_1$ to $i_2$ molecules in $L$ and $H$ reservoirs can be expressed as
\begin{align}
{{{c_{{i_1},L}}} \mathord{/} {{c_{{i_2},L}}}} = {{{c_{{i_1},H}}} \mathord{/} {{c_{{i_2},H}}}} = \frac{1}{{{c_2}}} - 1.
\end{align}

After moving molecules between the reservoirs, a concentration ratio difference emerges between them. A decision-making process is then proposed to analyze the transmitter's performance under various molecular movement strategies, considering the concentration differences caused by these strategies and a limited yet equivalent amount of input free energy. By examining the statistical characteristics of the transmitted molecules, the relationship between the error introduced by the transmitter and the associated energy cost is investigated to characterize the system's efficiency. Through a comparative analysis of the communication system's performance under different molecule movement strategies, an optimal molecule movement scheme is identified that maximizes communication efficiency within the constraints of limited energy consumption.

Given the symmetry in the initial conditions (e.g., the number and concentrations of molecules in the reservoirs), the performance of moving $i_1$ or $i_2$ molecules from $L$ to $H$ reservoir is equivalent to their movement in the opposite direction, from $H$ to $L$ reservoir. Therefore, we focus exclusively on the $L$ to $H$ reservoir direction to streamline the analysis.
\subsection{Moving $i_2$ molecules from $L$ to $H$ reservoir}

In this subsection, we consider move $m_2$ $i_2$ molecules from the $L$ reservoir to the $H$ reservoir, considering that ${m_2} \ll \{ {n_{L}},{n_{H}}\}$ , then, the concentrations of  $i_2$ molecules in the $L$ and $H$ reservoirs can be approximated as
\begin{align}
\label{c_{2,i_2}}
\left\{ 
    \begin{aligned}
{c_{{i_2},L}} &= \frac{{{n_{L}}{c_2} - {m_2}}}{{{n_{L}}}} = {c_2} - \frac{{{m_2}}}{{{n_{L}}}},\\
{c_{{i_2},H}} &= \frac{{{n_{H}}{c_2} + {m_2}}}{{{n_{H}}}} = {c_2} + \frac{{{m_2}}}{{{n_{H}}}}.
    \end{aligned} \right.
\end{align}

After moving $m_2$ $i_2$  molecules from the $L$ reservoir to the $H$ reservoir, the ratio of  $i_1$ to $i_2$ molecules in the $L$ reservoir increases, resulting in ${c_{{i_1},L}}\mathord / {c_{{i_2},L}} \ge \frac{1}{{{c_2}}} - 1$. Conversely, the ratio in the $H$ reservoir decreases, such that ${{c_{{i_1},H}} \mathord / {c_{{i_2},H}}} < \frac{1}{{{c_2}}} - 1$. Therefore, the decision criterion for the information carried by the molecular stream can be expressed as
\begin{align}
{r_{rx,i}} = \left\{ 
\begin{aligned}
&0,{{{c_{{i_1},L}}} \mathord{\left/
 {\vphantom {{{c_{{i_1},L}}} {{c_{{i_2},L}}}}} \right.
 \kern-\nulldelimiterspace} {{c_{{i_2},L}}}} \ge \frac{1}{{{c_2}}} - 1,\\
&1,{{{c_{{i_1},H}}} \mathord{\left/
 {\vphantom {{{c_{{i_1},H}}} {{c_{{i_2},H}}}}} \right.
 \kern-\nulldelimiterspace} {{c_{{i_2},H}}}} < \frac{1}{{{c_2}}} - 1.
\end{aligned} \right.
\end{align}

Assuming that to transmit bit 0, $N_m$ molecules are selected from the low reservoir, which satisfies ${{c_{{i_1},L}} \mathord / {c_{{i_2},L}}} \ge \frac{1}{{{c_2}}} - 1$. Additionally, for correct transmission, the minimum number of $i_1$ molecules emitted from the $L$ reservoir should be $\left\lfloor {N_m(1 - {c_2})} \right\rfloor + 1$. Therefore, the probability of correctly transmitting bit 0 from the low reservoir can be expressed as
 \begin{align}
{P_{{i_2}}}(Y = 0\left| {X = 0} \right.) = \frac{{\sum\limits_{i = \left\lfloor {{N_m}(1 - {c_2})} \right\rfloor + 1}^{N_m} {C_{{n_{L}}(1 - {c_{{i_2},L}})}^iC_{{n_{L}}{c_{{i_2},L}}}^{{N_m} - i}} }}{{C_{{n_{L}}}^{N_m}}}.
\end{align}

Similarly, for transmitting bit 1, the probability of correctly transmitting it from the high reservoir can be expressed as
\begin{align}
{P_{{i_2}}}(Y = 1\left| {X = 1} \right.) = \frac{{\sum\limits_{i = \left\lfloor {N_m{c_2}} \right\rfloor + 1}^{N_m} {C_{{n_{H}}{c_{{i_2},H}}}^iC_{{n_{H}}(1 - {c_{{i_2},H}})}^{N_m - i}} }}{{C_{{n_{H}}}^{N_m}}}.
\end{align}

Assuming an equal probability of transmitting bit 0 or bit 1, the error rate used to characterize the transmitter's performance when moving $i_2$ molecules can be expressed as
\begin{align}
\begin{split}
{P_{e,{i_2}}} &= P(X = 0) \times P(Y = 1\left| {X = 0} \right.)\\
&+P(X = 1) \times P(Y = 0\left| {X = 1} \right.)\\
&= \frac{1}{2}\Big[(1 - {P_{{i_2}}}(Y = 0\left| {X = 0} \right.))\\
&+(1 - {P_{{i_2}}}(Y = 1\left| {X = 1} \right.))\Big]\\
&= \frac{1}{2}\Bigg[2 - \frac{\sum\limits_{i = \left\lfloor {{N_m}(1 - {c_2})} \right\rfloor  + 1}^{N_m} {C_{{n_{L}}(1 - {c_{{i_2},L}})}^iC_{{n_{L}}{c_{{i_2},L}}}^{{N_m} - i}} }{C_{{n_{L}}}^{N_m}}\\
&- \frac{\sum\limits_{i = \left\lfloor {{N_m}{c_2}} \right\rfloor  + 1}^{N_m} {C_{{n_{H}}{c_{{i_2},H}}}^iC_{{n_{H}}(1 - {c_{{i_2},H}})}^{N_m - i}} }{C_{{n_{H}}}^{N_m}}\Bigg],
\end{split}
\end{align}
where $\frac{{\sum\limits_{i = \left\lfloor {N_m(1 - {c_2})} \right\rfloor  + 1}^{N_m} {C_{{n_{L}}(1 - {c_{{i_2},L}})}^iC_{{n_{L}}{c_{{i_2},L}}}^{N_m - i}} }}{{C_{{n_{L}}}^{N_m}}}$ follows a hypergeometric distribution, and as $n_{L}$ is large, then, the hypergeometric distribution can be approximated by a binomial distribution ${X_{{i_2},0}}(i) \sim B(i,1 - {c_{{i_2},L}})$. Considering that $N_m$ is also large enough, the binomial distribution can be further approximated by a normal distribution. Therefore, ${X_{{i_2},0}}(i) \sim N(\mu _0,\sigma _0^2)$, where
\begin{align}
\begin{split}
{\mu _0} &= N_m(1 - {c_{{i_2},L}}),\\
\sigma _0^2 &= N_m{c_{{i_2},L}}(1 - {c_{{i_2},L}}).
\end{split}
\end{align}

Therefore, the probability of transmitting 0 and correctly judging it can be approximated as
\begin{align}
\begin{split}
\label{P_{i_2,00}}
{P_{{i_2}}}(Y = 0\left| {X = 0} \right.) &= \Phi \Big(\frac{{N_m - {\mu _0}}}{{{\sigma _0}}}\Big) \\
&- \Phi \Big(\frac{{\left\lfloor {N_m(1 - {c_2})} \right\rfloor  + 1 - {\mu _0}}}{{{\sigma _0}}}\Big),
\end{split}
\end{align}
where $\Phi (x) = \frac{1}{{\sqrt {2\pi } }}\int_{ - \infty }^x {{e^{ - \frac{{{t^2}}}{2}}}} dt$ is the cumulative distribution function of the standard normal distribution.

In MC, it is reasonable to assume that $N_m({1-c_2})$ is large. Therefore, $\left\lfloor {N_m(1 - {c_2})} \right\rfloor +1 - {\mu _0}$ can be approximated as $N_m{(1-c_2)} - {\mu _0}$. Consequently, (\ref{P_{i_2,00}}) can be rewritten as
\begin{align}
{P_{{i_2}}}(Y = 0\left| {X = 0} \right.) = \Phi \Big(\frac{{N_m - {\mu _0}}}{{{\sigma _0}}}\Big) - \Phi \Big(\frac{{N_m(1 - {c_2}) - {\mu _0}}}{{{\sigma _0}}}\Big).
\end{align}

Similarly, 
% considering the large values of $n_{H}$ and $N_m$, 
the hypergeometric distribution $\frac{{\sum\limits_{i = \left\lfloor {N_m{c_2}} \right\rfloor  + 1}^{N_m} {C_{{n_{H}}{c_{{i_2},H}}}^iC_{{n_{H}}(1 - {c_{{i_2},H}})}^{N_m - i}} }}{{C_{{n_{H}}}^{N_m}}}$ can be further approximated by a normal distribution ${X_{{i_2},1}}(i) \sim N(\mu _1,\sigma _1^2)$, where
\begin{align}
\begin{split}
{\mu _1} &= N_m{c_{{i_2},H}},\\
\sigma _1^2 &= N_m{c_{{i_2},H}}(1 - {c_{{i_2},H}}).
\end{split}
\end{align}

Therefore, the probability of transmitting bit 1 and correctly judging it can be approximated as
\begin{align}
\label{P_{i_2,11}}
{P_{{i_2}}}(Y = 1\left| {X = 1} \right.) = \Phi \Big(\frac{{N_m - {\mu _1}}}{{{\sigma _1}}}\Big) - \Phi \Big(\frac{{\left\lfloor {N_m{c_2}} \right\rfloor  + 1 - {\mu _1}}}{{{\sigma _1}}}\Big).
\end{align}

Similarly, assuming $N_m{c_2}$ is large, $\left\lfloor {N_m{c_2}} \right\rfloor  + 1 - {\mu _1}$  can be approximated as $N_m{c_2} - {\mu _1}$. Therefore, equation (\ref{P_{i_2,11}}) can be rewritten as
\begin{align}
{P_{{i_2}}}(Y = 1\left| {X = 1} \right.) = \Phi (\frac{{N_m - {\mu _1}}}{{{\sigma _1}}}) - \Phi (\frac{{N_m{c_2} - {\mu _1}}}{{{\sigma _1}}}).
\end{align}

Therefore, when $m_2$ $i_2$ molecules are moved from the low reservoir to the high reservoir, the error rate of information transmission by the transmitter within a given time slot can be approximated as
\begin{align}
\begin{split}
\label{P_{{i_2}}}
{P_{e,{i_2}}} &= \frac{1}{2}\Bigg[2 - \Phi \Big(\frac{{N_m - {\mu _0}}}{{{\sigma _0}}}\Big) + \Phi \Big(\frac{{N_m(1 - {c_2}) - {\mu _0}}}{{{\sigma _0}}}\Big)\\
&- \Phi \Big(\frac{{N_m - {\mu _1}}}{{{\sigma _1}}}\Big) + \Phi \Big(\frac{{N_m{c_2} - {\mu _1}}}{{{\sigma _1}}}\Big)\Bigg]\\
&= \frac{1}{2}\Bigg[2 - \Phi \Big(\frac{{N_m{c_{{i_2},L}}}}{{\sqrt {N_m{c_{{i_2},L}}(1 - {c_{{i_2},L}})} }}\Big)\\
&+ \Phi \Big(\frac{{N_m({c_{{i_2},L}} - {c_2})}}{{\sqrt {N_m{c_{{i_2},L}}(1 - {c_{{i_2},L}})} }}\Big) - \Phi \Big(\frac{{N_m(1 - {c_{{i_2},H}})}}{{\sqrt {N_m{c_{{i_2},H}}(1 - {c_{{i_2},H}})} }}\Big)\\ 
&+ \Phi \Big(\frac{{N_m({c_2} - {c_{{i_2},H}})}}{{\sqrt {N_m{c_{{i_2},H}}(1 - {c_{{i_2},H}})} }}\Big)\Bigg].
\end{split}
\end{align}

Substituting equation (\ref{c_{2,i_2}}) in (\ref{P_{{i_2}}}), the  expression for the error rate when moving $m_2$ $i_2$ molecules from the low reservoir to the high reservoir can be obtained as
\begin{align}
\label{Pe_{i_2}}
\begin{split}
{P_{e,{i_2}}} &= \frac{1}{2}\Bigg[2 - \Phi \Big(\frac{{N_m({c_2} - \frac{{{m_2}}}{{{n_{L}}}})}}{{\sqrt {N_m({c_2} - \frac{{{m_2}}}{{{n_{L}}}})(1 - {c_2} + \frac{{{m_2}}}{{{n_{L}}}})} }}\Big)\\ 
&+ \Phi \Big(\frac{{ - N_m\frac{{{m_2}}}{{{n_{L}}}}}}{{\sqrt {N_m({c_2} - \frac{{{m_2}}}{{{n_{L}}}})(1 - {c_2} + \frac{{{m_2}}}{{{n_{L}}}})} }}\Big)\\
&- \Phi \Big(\frac{{N_m(1 - {c_2} - \frac{{{m_2}}}{{{n_{H}}}})}}{{\sqrt {N_m({c_2} + \frac{{{m_2}}}{{{n_{H}}}})(1 - {c_2} - \frac{{{m_2}}}{{{n_{H}}}})} }}\Big)\\ 
&+ \Phi \Big(\frac{{ - N_m\frac{{{m_2}}}{{{n_{H}}}}}}{{\sqrt {N_m({c_2} + \frac{{{m_2}}}{{{n_{H}}}})(1 - {c_2} - \frac{{{m_2}}}{{{n_{H}}}})} }}\Big)\Bigg],
\end{split}
\end{align}
where 
% $N_m$ is the number of transmitted molecules for per bit, $c_2$ is the initial concentration of $i_2$ molecules, and ${m_2}$ is the number of moved molecules from the low reservoir to the high reservoir, 
${m_2} = \sqrt {\frac{{{c_2}{n}}}{{2KT}}E}$.

\subsection{Moving $i_1$ molecules from $L$ to $H$ reservoir}
In this subsection, we consider the movement of $i_1$ molecules from the $L$ reservoir to the $H$ reservoir. When $m_1$ $i_1$ molecules are moved from the $L$ reservoir to the $H$ reservoir, given that ${m_1} \ll \{{n_{L}},{n_{H}}\}$, the concentrations of $i_2$ molecules in the $L$ and $H$ reservoirs can be approximated as
% \begin{align}
% \label{c_{2,i_1}}
% \left\{ 
% \begin{aligned}
% {c_{{i_2},L}} &= 1 - {c_{{i_1},L}} = 1 - \frac{{{n_{L}}(1 - {c_2}) - {m_1}}}{{{n_{L}}}} = {c_2} + \frac{{{m_1}}}{{{n_{L}}}},\\
% {c_{{i_2},H}} &= 1 - {c_{{i_1},H}} = 1 - \frac{{{n_{H}}(1 - {c_2}) + {m_1}}}{{{n_{H}}}} = {c_2} - \frac{{{m_1}}}{{{n_{H}}}}.
% \end{aligned}\right.
% \end{align}
\begin{align}
\label{c_{2,i_1}}
\left\{ 
\begin{aligned}
{c_{{i_2},L}} &= 1 - {c_{{i_1},L}} = {c_2} + \frac{{{m_1}}}{{{n_{L}}}},\\
{c_{{i_2},H}} &= 1 - {c_{{i_1},H}} =  {c_2} - \frac{{{m_1}}}{{{n_{H}}}}.
\end{aligned}\right.
\end{align}

After moving $m_1$ $i_1$ molecules from the $L$ reservoir to the $H$ reservoir, the ratio of $i_1$ to $i_2$ molecules in the $L$ reservoir decreases,  resulting in ${c_{{i_1},L}} \mathord{/c_{{i_2},L}} \le \frac{1}{{{c_2}}} - 1$, while the ratio in the $H$ reservoir increases, such that  ${c_{{i_1},H}} \mathord{/c_{{i_2},H}} > \frac{1}{c_2} - 1$.  Consequently, the decision criterion for the information carried by the molecular stream can be expressed as
\begin{align}
\begin{split}
{r_{rx,i}} = \left\{ 
\begin{aligned}
&0,{{{c_{{i_1},L}}} \mathord{/} {{c_{{i_2},L}}}} \le \frac{1}{{{c_2}}} - 1,\\
&1,{{{c_{{i_1},H}}} \mathord{/} {{c_{{i_2},H}}}} > \frac{1}{{{c_2}}} - 1.
\end{aligned} \right.
\end{split}
\end{align}

 Assuming $N_m$  molecules are emitted by the transmitter for each bit. To transmit bit 0, $N_m$ molecules are selected from the low reservoir, which should satisfy ${c_{{i_1},L}} \mathord{/c_{{i_2},L}} \le \frac{1}{{c_2}} - 1$. Additionally, the minimum number of $i_2$ molecules emitted from the $L$ reservoir should be $\left\lfloor {{N_m}{c_2}} \right\rfloor  + 1$. Therefore, the probability of correctly transmitting bit 0 from the low reservoir can be expressed as
 \begin{align}
{P_{{i_1}}}(Y = 0\left| {X = 0} \right.) = \frac{{\sum\limits_{i = \left\lfloor {{N_m}{c_2}} \right\rfloor  + 1}^{N_m} {C_{{n_{L}}{c_{{i_2},L}}}^iC_{{n_{L}}(1 - {c_{{i_2},L}})}^{{N_m} - i}} }}{{C_{{n_{L}}}^{N_m}}}.
 \end{align}
 
Similarly, for transmitting bit 1, the probability of correctly transmitting the molecules from the high reservoir can be expressed as
\begin{align}
{P_{{i_1}}}(Y = 1\left| {X = 1} \right.) = \frac{{\sum\limits_{i = \left\lfloor {{N_m}(1 - {c_2})} \right\rfloor  + 1}^{N_m} {C_{{n_{H}}(1 - {c_{{i_2},H}})}^iC_{{n_{H}}{c_{{i_2},H}}}^{{N_m} - i}} }}{{C_{{n_{H}}}^{N_m}}}.
\end{align}

As discussed before, assuming that the probabilities of transmitting bit 0 or bit 1 are both $1/2$, the error rate used to characterize the transmitter's performance when moving $i_1$ molecules can be expressed as
\begin{align}
\begin{split}
{P_{e,{i_1}}} &= P(X = 0) \cdot P(Y = 1\left| {X = 0} \right.)\\
&+ P(X = 1) \cdot P(Y = 0\left| {X = 1} \right.)\\
&= \frac{1}{2}[(1 - {P_{{i_1}}}(Y = 0\left| {X = 0} \right.))\\
&+ (1 - {P_{{i_1}}}(Y = 1\left| {X = 1} \right.))]\\
&= \frac{1}{2}\Big(2 - \frac{{\sum\limits_{i = \left\lfloor {{N_m}{c_2}} \right\rfloor  + 1}^{N_m} {C_{{n_{L}}{c_{{i_2},L}}}^iC_{{n_{L}}(1 - {c_{{i_2},L}})}^{{N_m} - i}} }}{{C_{{n_{L}}}^{N_m}}}\\
&- \frac{{\sum\limits_{i = \left\lfloor {{N_m}(1 - {c_2})} \right\rfloor  + 1}^{N_m} {C_{{n_{H}}(1 - {c_{{i_2},H}})}^iC_{{n_{H}}{c_{{i_2},H}}}^{{N_m} - i}} }}{{C_{{n_{H}}}^{N_m}}}\Big),
\end{split}
\end{align}
where $\frac{{\sum\limits_{i = \left\lfloor {{N_m}{c_2}} \right\rfloor + 1}^{N_m} {C_{{n_{L}}{c_{{i_2},L}}}^iC_{{n_{L}}(1 - {c_{{i_2},L}})}^{{N_m} - i}} }}{{C_{{n_{L}}}^{N_m}}}$ follows a hypergeometric distribution, and as $n_{L}$ and $N_m$ are sufficiently large, 
% then, the hypergeometric distribution can be approximated by a binomial distribution ${X_{{i_1},0}}(i) \sim B(i,{c_{{i_2},L}})$. Considering that ${N_m}$ is also large enough, the binomial distribution can be further 
it can be approximated by a normal distribution, ${X_{{i_1},0}} \sim N({\mu _0},\sigma _0^2)$, where
\begin{align}
\begin{split}
{\mu _0} &= {N_m}{c_{{i_2},L}}\\
\sigma _0^2 &= {N_m}{c_{{i_2},L}}(1 - {c_{{i_2},L}}).
\end{split}
\end{align}

Similarly to the derivation from (\ref{P_{i_2,00}}) - (14), when $m_1$ $i_1$ molecules are moved from the low reservoir to the high reservoir, the error rate of information transmission by the transmitter in a given time slot can be approximated as
\begin{align}
\begin{split}
\label{Pe_{i_1_1}}
{P_{e,{i_1}}} &= \frac{1}{2}\Bigg[2 - \Phi (\frac{{{N_m} - {\mu _0}}}{{{\sigma _0}}}) + \Phi (\frac{{{N_m}{c_2} - {\mu _0}}}{{{\sigma _0}}})\\
&- \Phi (\frac{{{N_m} - {\mu _1}}}{{{\sigma _1}}}) + \Phi (\frac{{{N_m}(1 - {c_2}) - {\mu _1}}}{{{\sigma _1}}})\Bigg]\\
&= \frac{1}{2}\Bigg[2 - \Phi \Big(\frac{{{N_m}(1 - {c_{{i_2},L}})}}{{\sqrt {{N_m}{c_{{i_2},L}}(1 - {c_{{i_2},L}})} }}\Big) \\
&+ \Phi \Big(\frac{{{N_m}({c_2} - {c_{{i_2},L}})}}{{\sqrt {{N_m}{c_{{i_2},L}}(1 - {c_{{i_2},L}})} }}\Big)\\
&- \Phi \Big(\frac{{{N_m}{c_{{i_2},H}}}}{{\sqrt {{N_m}{c_{{i_2},H}}(1 - {c_{{i_2},H}})} }}\Big) \\
&+ \Phi \Big(\frac{{{N_m}({c_{{i_2},H}} - {c_2})}}{{\sqrt {{N_m}{c_{{i_2},H}}(1 - {c_{{i_2},H}})} }}\Big)\Bigg].
\end{split}
\end{align}

Substituting equation (\ref{c_{2,i_1}}) in (\ref{Pe_{i_1_1}}), the expression for the error rate when moving $m_1$ $i_1$ molecules from the low reservoir to the high reservoir can be obtained as
\begin{align}
\label{Pe_{i_1}}
\begin{split}
{P_{e,{i_1}}} &= \frac{1}{2}\Bigg[2 - \Phi \Big(\frac{{{N_m}(1 - {c_2} - \frac{{{m_1}}}{{{n_{L}}}})}}{{\sqrt {{N_m}({c_2} + \frac{{{m_1}}}{{{n_{L}}}})(1 - {c_2} - \frac{{{m_1}}}{{{n_{L}}}})} }}\Big)\\
&+ \Phi \Big(\frac{{ - {N_m}\frac{{{m_1}}}{{{n_{L}}}}}}{{\sqrt {{N_m}({c_2} + \frac{{{m_1}}}{{{n_{L}}}})(1 - {c_2} - \frac{{{m_1}}}{{{n_{L}}}})} }}\Big)\\
&- \Phi \Big(\frac{{{N_m}({c_2} - \frac{{{m_1}}}{{{n_{H}}}})}}{{\sqrt {{N_m}({c_2} - \frac{{{m_1}}}{{{n_{H}}}})(1 - {c_2} + \frac{{{m_1}}}{{{n_{H}}}})} }}\Big)\\
&+ \Phi \Big(\frac{{ - {N_m}\frac{{{m_1}}}{{{n_{H}}}}}}{{\sqrt {{N_m}({c_2} - \frac{{{m_1}}}{{{n_{H}}}})(1 - {c_2} + \frac{{{m_1}}}{{{n_{H}}}})} }}\Big)\Bigg].
\end{split}
\end{align}
where 
% $c_2$ is the initial concentration of $i_2$ molecules in the reservoirs and $m_1$ is the number of moved molecules from the low reservoir to the high reservoir,
${m_1} = \sqrt {\frac{{(1 - {c_2}){n}}}{{2KT}}E}$.

From equation (\ref{Pe_{i_2}}) and equation (\ref{Pe_{i_1}}), it is shown that under the same conditions, ${P_{e,{i_1}}}$ and ${P_{e,{i_2}}}$ can be considered as univariate functions of $c_2$. It can be observed that ${P_{e,{i_1}}}(1 - {c_2}) = {P_{e,{i_2}}}{c_2}$, indicating that ${P_{e,{i_1}}}$ and ${P_{e,{i_2}}}$ are symmetric with respect to ${c_2} = 0.5$.
This demonstrates that when the initial concentrations of the two molecules are the same, the bit error rate is identical regardless of which molecule is moved.

Furthermore, with the same limited free energy input, moving the higher-concentration molecules results in improved transmitter performance.

\subsection{How the number of molecules per move affects  transmitter performance}
In this subsection, we analyze the transmitter performance based on the number of molecules transferred per operation.  As outlined in subsections A and B, the initial concentrations of $i_1$ and $i_2$ molecules differ, with the initial fraction of $i_2$  molecules being larger than that of $i_1$  molecules.

To evaluate the performance in terms of the number of molecules transferred per operation, we begin by assuming that equilibrium is established between the two types of molecules in the low reservoir after the transfer process. Subsequently, $i_2$ molecules are moved from the low reservoir to the high reservoir, subject to limited energy input. The total number of moved $i_2$ molecules is given by $m = (2c - 1)n_L$, where $n_L$ represents the total number of molecules in the low reservoir, $c$ is the initial concentration of $i_2$ molecules. If $m_0$ molecules are moved per transfer, the energy consumed per transfer can then be expressed as
\begin{align}
{E_i} = \frac{{2KT m_0^2}}{n} \frac{1}{{c - \frac{{i{m_0}}}{{{n_L}}}}} = \frac{{KT m_0^2}}{{c{n_L} - i{m_0}}},i = 0,1,...,\frac{m}{m_0} - 1.
\end{align}

% First, we analyze the effect of the number of per move molecules $m_0$ on the energy cost $E_i$. The number of moved steps $i=m/m_0-1$. Then, the partial derivative of $im_0$ with respect to $m_0$ can be expressed as
% \begin{align}
%     \frac{{\partial (i{m_0})}}{{\partial {m_0}}} =  - \frac{m}{{m_0^2}} \cdot {m_0} + (\frac{m}{{{m_0}}} - 1) \cdot 1 =  - 1
% \end{align}

Considering the number of moved molecules, $m=(2c-1)n_L$, the partial derivative of $E_i$ with respect to $m_0$ can be expressed as
\begin{align}
\begin{split}
\frac{{\partial {E_{\rm{i}}}}}{{\partial {m_0}}} &= KT \cdot \frac{{2{m_0}(c{n_L} - i{m_0}) - m_0^2( - i)}}{{{{(c{n_L} - i{m_0})}^2}}}\\ 
&= KT \cdot \frac{{2{m_0}(c{n_L} - i{m_0}) + im_0^2}}{{{{(c{n_L} - i{m_0})}^2}}} > 0
\end{split}
\end{align}

Since the total number of transfers is $\frac{m}{{{m_0}}} - 1$, the total energy consumed can be expressed as
\begin{align}
\begin{split}
{E_{total}} &= \sum\limits_{i = 0}^{m/{m_0} - 1} {{E_i}}  = \sum\limits_{i = 0}^{m/{m_0} - 1} {\frac{{KT m_0^2}}{{c{n_L} - i{m_0}}}}.
\end{split}
\end{align}

To analyze the effect of the number of moved molecules per transfer, $m_0$, the partial derivative of the total energy consumed, $E_{total}$, with respect to $m_0$ can be expressed as
\begin{align}
\begin{split}
\frac{{\partial {E_{total}}}}{{\partial {m_0}}} = \sum\limits_{i = 0}^{m/{m_0} - 1} {\frac{{\partial {E_i}}}{{\partial {m_0}}}}  > 0.
\end{split}
\end{align}

Since the number of molecules moved per transfer, $m_0$, is much smaller than the total number of molecules in the reservoir, it follows that $\frac{{\partial {E_{total}}}}{{\partial {m_0}}} > 0$. This indicates that the total energy consumed, $E_{total}$, increases as the number of molecules moved per transfer, $m_0$, increases.

Therefore, for a given energy cost, moving a smaller number of molecules per transfer, $m_0$, results in a greater total number of molecules transferred.  Consequently, better transmitter performance is achieved.

\subsection{Movement strategies following molecule equilibrium in the low reservoir}
Based on the analysis in subsections $A$ through $C$, and under the assumption that the initial fraction of $i_2$ molecules is larger than that of $i_1$ molecules, a state can be achieved where the concentrations of $i_1$ and $i_2$ molecules are balanced in the low reservoir after moving $i_2$ molecules from the low reservoir to the high reservoir. At this point,
% considering the concentration of $i_2$ molecules in the low reservoir is $c$ and the concentration of $i_1$ molecules in the high reservoir is $1-c$. Then, 
the concentration of $i_2$ molecules in the low reservoir and $i_1$ molecules in the high reservoir can be expressed as
\begin{align}
\begin{split}
{c_{i_2, L}} &= c - \frac{m}{n_L},\\
{c_{i_1, H}} &= 1 - c - \frac{m}{n_H}.\\    
\end{split}
\end{align}

% where $c$ is the initial fraction of $i_2$ molecules.
Considering there is remaining free energy after moving $i_2$ molecules from the low reservoir to the high reservoir, and the concentrations of $i_1$ and $i_2$ molecules are balanced in the low reservoir, this energy can still be employed to move molecules between the reservoirs to improve transmitter performance. Then, three movement strategies between the reservoirs are considered: \\
(1) Continue moving $i_2$ molecules from the low reservoir to the high reservoir. \\
(2) Cross-move $i_2$ molecules from the low reservoir to the high reservoir while moving $i_1$ molecules from the high reservoir to the low reservoir. \\
(3) Continue moving $i_1$ molecules from the high reservoir to the low reservoir.

In the following, we analyze the energy cost and transmitter performance for each of these three movement strategies. Based on the analysis in subsection $C$, better transmitter performance can be achieved with a smaller number of molecules per transfer. Therefore, in this subsection, we consider the case where the number of molecules moved per transfer is $m_0 = 1$, meaning only one molecule is moved at a time.

(1) For the continuous movement of $i_2$ molecules from the low reservoir to the high reservoir, the energy consumed can be expressed as
\begin{align}
\begin{split}
{E_{i_2}} &= \frac{{2KT{{m}}_0^2}}{n}\Bigg[\frac{1}{{{c_{{i_2},L}}}} + \frac{1}{{{c_{{i_2},L}} - \frac{{{m_0}}}{{{n_L}}}}} + \frac{1}{{{c_{{i_2},L}} - \frac{{2{m_0}}}{{{n_L}}}}} + ... \\
&+ \frac{1}{{{c_{{i_2},L}} - \frac{{({i_{i_2}} - 1){m_0}}}{{{n_L}}}}}\Bigg], 
\end{split}
\end{align}
where $i_{i_2}$ is the number of transfers until the energy is exhausted during the continuous movement of $i_2$ molecules. Then, the concentration of $i_2$ molecules in the reservoirs can be expressed as
\begin{align}
{{c_{{i_2},L,1}} = c - \frac{m}{{{n_L}}} - \frac{{{i_{i_2}}{m_0}}}{{{n_L}}} = c - \frac{{m + {i_{i_2}}}}{{{n_L}}}},   
\end{align}
\begin{align}
{{c_{{i_2},H,1}} = c + \frac{m}{{{n_H}}} + \frac{{{i_{i_2}}{m_0}}}{{{n_H}}} = c + \frac{{m + {i_{i_2}}}}{{{n_H}}}},
\end{align}
where $m_0=1$ and $m=(2c-1)n_L$.

(2) For the cross-movement of $i_2$ molecules from the low reservoir to the high reservoir, while simultaneously moving $i_1$ molecules from the high reservoir to the low reservoir, the energy consumed can be expressed as
% \begin{align}
% \renewcommand{\arraystretch}{1.5}
% {E_{{i_1}{i_2}}} = \left\{ {\begin{array}{*{20}{c}}
% {\frac{{2KT{{m}}_0^2}}{n}\Big(\frac{1}{{{c_{{i_2},L}}}} + \frac{1}{{{c_{{i_1},H}} - \frac{{{m_0}}}{{{n_H}}}}} + \frac{1}{{{c_{{i_2},L}} - \frac{{2{m_0}}}{{{n_L}}}}} + ... \\
% + \frac{1}{{{c_{{i_2},L}} - \frac{{({i_{{i_1}{i_2}}} - 1){m_0}}}{{{n_L}}}}}\Big),{{{i}}_{{i_1}{i_2}}} = 1,3, \cdots, 2n - 1},\\
% {\frac{{2KT{{m}}_0^2}}{n}\Big(\frac{1}{{{c_{{i_2},L}}}} + \frac{1}{{{c_{{i_1},H}} - \frac{{{m_0}}}{{{n_H}}}}} + \frac{1}{{{c_{{i_2},L}} - \frac{{2{m_0}}}{{{n_L}}}}} + ...\\
% + \frac{1}{{{c_{{i_1},H}} - \frac{{({i_{{i_1}{i_2}}} - 1){m_0}}}{{{n_H}}}}}\Big),{{{i}}_{{i_1}{i_2}}} = 2,4, \cdots, 2n},
% \end{array}} \right.
% \end{align}
\begin{align}
\renewcommand{\arraystretch}{1.5}
E_{i_1 i_2} = \left\{ 
\begin{array}{*{20}{c}}
    \frac{2KTm_0^2}{n} \Big( \frac{1}{c_{i_2,L}} + \frac{1}{c_{i_1,H} - \frac{m_0}{n_H}} + \frac{1}{c_{i_2,L} - \frac{2 m_0}{n_L}} + ... \\
    + \frac{1}{c_{i_2, L} - \frac{(i_{i_1 i_2} - 1)m_0}{n_L}} \Big), i_{i_1 i_2} = 1,3, \cdots, 2n - 1,
    \\
    \frac{2KTm_0^2}{n} \Big( \frac{1}{c_{i_2,L}} + \frac{1}{c_{i_1,H} - \frac{m_0}{n_H}} + \frac{1}{c_{i_2,L} - \frac{2 m_0}{n_L}} + ... \\
    + \frac{1}{c_{i_2, H} - \frac{(i_{i_1 i_2} - 1)m_0}{n_H}} \Big), i_{i_1 i_2} = 2,4, \cdots, 2n,    
    % {\frac{{2KT{{m}}_0^2}}{n}\Big(\frac{1}{{{c_{{i_2},L}}}} + \frac{1}{{{c_{{i_1},H}} - \frac{{{m_0}}}{{{n_H}}}}} + \frac{1}{{{c_{{i_2},L}} - \frac{{2{m_0}}}{{{n_L}}}}} + ...\\
    % + \frac{1}{{{c_{{i_1},H}} - \frac{{({i_{{i_1}{i_2}}} - 1){m_0}}}{{{n_H}}}}}\Big),{{{i}}_{{i_1}{i_2}}} = 2,4, \cdots, 2n},
\end{array} \right.
\end{align}
where $i_{{i_1}{i_2}}$ is the number of transfers until the energy is exhausted during the cross-movement of $i_2$ molecules from the low reservoir to the high reservoir, while simultaneously moving $i_1$ molecules from the high reservoir to the low reservoir. Then, the concentrations of $i_2$ molecules in the reservoirs can be expressed as
\begin{align}
{c_{{i_2},L,2}} = c - \frac{m}{{{n_L}}} - \frac{{{i_{{i_1}{i_2}}}{m_0}}}{{{n_L}}} = c - \frac{{m + {i_{{i_1}{i_2}}}}}{{{n_L}}},
\end{align}
\begin{align}
{c_{{i_2},H,2}} = c + \frac{m}{{{n_H}}} + \frac{{{i_{{i_1}{i_2}}}{m_0}}}{{{n_H}}} = c + \frac{{m + {i_{{i_1}{i_2}}}}}{n_H}.
\end{align}

(3) For the movement of $i_1$ molecules from the high reservoir to the low reservoir, the energy consumed can be expressed as
\begin{align}
\begin{split}
E_{i_1} &= \frac{2KTm_0^2}{n} \left[ \frac{1}{c_{{i_1},H}} + \frac{1}{c_{{i_1},H} - \frac{m_0}{n_H}} 
+ \frac{1}{c_{{i_1},H} - \frac{2m_0}{n_H}} \right. \\
&\quad \left. + \ldots + \frac{1}{c_{{i_1},H} - \frac{(i_{i_1} - 1)m_0}{n_H}} \right],
\end{split}
\end{align}
where $i_{i_1}$ is the number of transfers until the energy is exhausted during the movement of $i_1$ molecules from the high reservoir to the low reservoir. Then, the concentration of ${i_2}$ molecules in the reservoirs can be expressed as
\begin{align}
{c_{{i_2},L,3}} = c - \frac{m}{{{n_L}}} - \frac{{{i_{i_1}}{m_0}}}{{{n_H}}} = c - \frac{{m + {i_{i_1}}}}{{{n_L}}},
\end{align}
\begin{align}
{c_{{i_2},H,3}} = c + \frac{m}{{{n_H}}} + \frac{{{i_{i_1}}{m_0}}}{{{n_H}}} = c + \frac{{m + {i_{i_1}}}}{{{n_H}}}.
\end{align}

As $c_{i_2,L}=c-m/n_L$, $c_{i_1,H}=1-c-m/n_H$, and $c>(1-c)$, then,
% there is $c_{{i_2},L}>c_{{i_1},H}$, it follows that
$\frac{1}{{{c_{{i_2},L}}}} < \frac{1}{{{c_{{i_1},H}}}}$. Therefore, for the $k$th movement, $\frac{1}{{{c_{{i_2},L}} - \frac{{k{m_0}}}{{{n_L}}}}} < \frac{1}{{{c_{{i_1},H}} - \frac{{k{m_0}}}{{{n_H}}}}}$. As a result, the continuous movement of $i_2$ molecules from the low reservoir to the high reservoir leads to the largest number of molecules transferred. Cross-move $i_2$ molecules from the low reservoir to the high reservoir while simultaneously moving $i_1$ molecules from the high reservoir to the low reservoir transfers more molecules. Finally, moving $i_1$ molecules from the high reservoir to the low reservoir results in the fewest molecules transferred,
namely $i_{i_2}>i_{{i_1}{i_2}}>i_{i_1}$. Therefore, $c_{{i_2},L,1}<c_{{i_2},L,2}<c_{{i_2},L,3}$ and $c_{{i_2},H,1}>c_{{i_2},H,2}>c_{{i_2},H,3}$. In all three movement strategies, the concentration of $i_2$ molecules decreases in the low reservoir while increasing in the high reservoir. Therefore, based on (\ref{P_{{i_2}}}), the BER raised from the transmitter can be expressed as 
\begin{align}
\begin{split}
{P_e} &= \frac{1}{2}\Bigg[2 - \Phi \Big(\frac{{{N_m}{c_{{i_2},L}}}}{{\sqrt {{N_m}{c_{{i_2},L}}(1 - {c_{{i_2},L}})} }}\Big) \\
&\qquad+ \Phi \Big(\frac{{{N_m}({c_{{i_2},L}} - c)}}{{\sqrt {{N_m}{c_{{i_2},L}}(1 - {c_{{i_2},L}})} }}\Big)\\
&\qquad- \Phi \Big(\frac{{{N_m}(1 - {c_{{i_2},H}})}}{{\sqrt {{N_m}{c_{{i_2},H}}(1 - {c_{{i_2},H}})} }}\Big) \\
&\qquad+ \Phi \Big(\frac{{{N_m}(c - {c_{{i_2},H}})}}{{\sqrt {{N_m}{c_{{i_2},H}}(1 - {c_{{i_2},H}})} }}\Big)\Bigg].
\end{split}
\end{align}
Assuming $A = \frac{{{N_m}{c_{{i_2},L}}}}{{\sqrt {{N_m}{c_{{i_2},L}}(1 - {c_{{i_2},L}})} }}$, $B = \frac{{{N_m}({c_{{i_2},L}} - c)}}{{\sqrt {{N_m}{c_{{i_2},L}}(1 - {c_{{i_2},L}})} }}$, $C = \frac{{{N_m}(1 - {c_{{i_2},H}})}}{{\sqrt {{N_m}{c_{{i_2},H}}(1 - {c_{{i_2},H}})} }}$, and $D = \frac{{{N_m}(c - {c_{{i_2},H}})}}{{\sqrt {{N_m}{c_{{i_2},H}}(1 - {c_{{i_2},H}})} }}$.
Given that these three movement strategies are based on balanced concentrations in the low reservoir, the conditions $0<c_{{i_2},L}<0.5$ and $0.5<c_{{i_2},H}<1$ hold. Additionally, $c_{{i_2},L}<c$ and $c_{{i_2},H}>c$. Consequently, $A>0$, $B<0$, $C>0$, and $D<0$ apply. Then,
\begin{align}
{\frac{{\partial \Phi }}{{\partial A}} = \frac{1}{{\sqrt {2\pi } }}\exp ( - {A^2}/2) \cdot ( - A) < 0},
\end{align}
\begin{align}
{\frac{{\partial \Phi }}{{\partial B}} = \frac{1}{{\sqrt {2\pi } }}\exp ( - {B^2}/2) \cdot ( - B) > 0},
\end{align}
\begin{align}
{\frac{{\partial \Phi }}{{\partial C}} = \frac{1}{{\sqrt {2\pi } }}\exp ( - {C^2}/2) \cdot ( - C) < 0},
\end{align}
\begin{align}
{\frac{{\partial \Phi }}{{\partial D}} = \frac{1}{{\sqrt {2\pi } }}\exp ( - {D^2}/2) \cdot ( - D) > 0}.
\end{align}

Meanwhile,
\begin{align}
\begin{split}
\frac{{\partial A}}{{\partial {c_{{i_2},L}}}} &= \frac{{\partial \Big(\sqrt {\frac{{{N_m}{c_{{i_2},L}}}}{{1 - {c_{{i_2},L}}}}}\Big)}}{{\partial {c_{{i_2},L}}}} \\
&= \frac{1}{{2A}} \frac{{{N_m}(1 - {c_{{i_2},L}}) - {N_m}{c_{{i_2},L}}(- 1)}}{{{{(1 - {c_{{i_2},L}})}^2}}} \\
&= \frac{{N_m}}{{2A{{(1 - {c_{{i_2},L}})}^2}}} > 0 ,  
\end{split}
\end{align}

\begin{align}
\begin{split}
\frac{{\partial B}}{{\partial {c_{{i_2},L}}}} &= \frac{{\partial \Big(\frac{{{N_m}({c_{{i_2},L}} - c)}}{{\sqrt {{N_m}{c_{{i_2},L}}(1 - {c_{{i_2},L}})} }}\Big)}}{{\partial {c_{{i_2},L}}}}\\
&= \frac{{{N_m}\sqrt {{N_m}{c_{{i_2},L}}(1 - {c_{{i_2},L}})}}}{{{N_m}{c_{{i_2},L}}(1 - {c_{{i_2},L}})}}\\
&- \frac{{{N_m}({c_{{i_2},L}} - c) \frac{{\sqrt {{N_m}{c_{{i_2},L}}(1 - {c_{{i_2},L}})} }}{{2{N_m}{c_{{i_2},L}}(1 - {c_{{i_2},L}})}} {N_m}(1 - 2{c_{{i_2},L}})}}{{{N_m}{c_{{i_2},L}}(1 - {c_{{i_2},L}})}}\\
% &= \frac{{M\sqrt {M{c_{L,{i_2}}}(1 - {c_{L,{i_2}}})}  - M({c_{L,{i_2}}} - c) \cdot \frac{{\sqrt {M{c_{L,{i_2}}}(1 - {c_{L,{i_2}}})} }}{{2M{c_{L,{i_2}}}(1 - {c_{L,{i_2}}})}} M(1 - 2{c_{L,{i_2}}})}}{{M{c_{L,{i_2}}}(1 - {c_{L,{i_2}}})}}\\
% &= \frac{{\sqrt {{N_m}{c_{L,{i_2}}}(1 - {c_{L,{i_2}}})}  - ({c_{L,{i_2}}} - c) (1 - 2{c_{L,{i_2}}})\frac{{\sqrt {{N_m}{c_{L,{i_2}}}(1 - {c_{L,{i_2}}})} }}{{2{c_{L,{i_2}}}(1 - {c_{L,{i_2}}})}}}}{{{c_{L,{i_2}}}(1 - {c_{L,{i_2}}})}}\\
% &= \frac{{\sqrt {{N_m}{c_{L,{i_2}}}(1 - {c_{L,{i_2}}})}  - ({c_{L,{i_2}}} - c) (1 - 2{c_{L,{i_2}}})\frac{{\sqrt {{N_m}{c_{L,{i_2}}}(1 - {c_{L,{i_2}}})} }}{{2{c_{L,{i_2}}}(1 - {c_{L,{i_2}}})}}}}{{{c_{L,{i_2}}}(1 - {c_{L,{i_2}}})}}\\
&= \frac{{\sqrt {{N_m}{c_{{i_2},L}}(1 - {c_{{i_2},L}})}}}{{{c_{{i_2},L}}(1 - {c_{{i_2},L}})}}\\
&+ \frac{{(c - {c_{{i_2},L}}) (1 - 2{c_{{i_2},L}})\frac{{\sqrt {{N_m}{c_{{i_2},L}}(1 - {c_{{i_2},L}})} }}{{2{c_{{i_2},L}}(1 - {c_{{i_2},L}})}}}}{{{c_{{i_2},L}}(1 - {c_{{i_2},L}})}} > 0,\\
% &= \frac{{\sqrt {{N_m}{c_{L,{i_2}}}(1 - {c_{L,{i_2}}})}  + (c - {c_{L,{i_2}}}) (1 - 2{c_{L,{i_2}}})\frac{{\sqrt {{N_m}{c_{L,{i_2}}}(1 - {c_{L,{i_2}}})} }}{{2{c_{L,{i_2}}}(1 - {c_{L,{i_2}}})}}}}{{{c_{L,{i_2}}}(1 - {c_{L,{i_2}}})}} > 0\\
\end{split}
\end{align}

\begin{align}
\begin{split}
\frac{{\partial C}}{{\partial {c_{{i_2},H}}}} &= \frac{{\partial \Big(\sqrt {\frac{{{N_m}(1 - {c_{{i_2},H}})}}{{{c_{{i_2},H}}}}} \Big)}}{{\partial {c_{{i_2},H}}}} \\
&= \frac{1}{{2C}} \frac{{ - {N_m}{c_{{i_2},H}} - {N_m}(1 - {c_{{i_2},H}})}}{{{c_{{i_2},H}^2}}} \\
&= \frac{{ - {N_m}}}{{2C {c_{{i_2},H}^2}}} < 0,
\end{split}
\end{align}
and 
\begin{align}
\begin{split}
\frac{{\partial D}}{{\partial {c_{{i_2},H}}}} &= \frac{{\partial \Big(\frac{{{N_m}(c - {c_{{i_2},H}})}}{{\sqrt {{N_m}{c_{{i_2},H}}(1 - {c_{{i_2},H}})} }}\Big)}}{{\partial {c_{{i_2},H}}}}\\
&= \frac{{ - {N_m}\sqrt {{N_m}{c_{{i_2},H}}(1 - {c_{{i_2},H}})}}}{{{N_m}{c_{{i_2},H}}(1 - {c_{{i_2},H}})}}\\
&- \frac{{{N_m}(c - {c_{{i_2},H}}) \frac{{\sqrt {{N_m}{c_{{i_2},H}}(1 - {c_{{i_2},H}})} }}{{2{N_m}{c_{{i_2},H}}(1 - {c_{{i_2},H}})}} {N_m}(1 - 2{c_{{i_2},H}})}}{{{N_m}{c_{{i_2},H}}(1 - {c_{{i_2},H}})}}\\
% &= \frac{{ - M\sqrt {M{c_{H,{i_2}}}(1 - {c_{H,{i_2}}})}  - M(c - {c_{H,{i_2}}}) \frac{{\sqrt {M{c_{H,{i_2}}}(1 - {c_{H,{i_2}}})} }}{{2M{c_{H,{i_2}}}(1 - {c_{H,{i_2}}})}} M(1 - 2{c_{H,{i_2}}})}}{{M{c_{H,{i_2}}}(1 - {c_{H,{i_2}}})}}\\
&= \frac{{ - \sqrt {{N_m}{c_{{i_2},H}}(1 - {c_{{i_2},H}})}}}{{{c_{{i_2},H}}(1 - {c_{{i_2},H}})}}\\
&- \frac{{({c_{{i_2},H}} - c) (2{c_{{i_2},H}} - 1)\frac{{\sqrt {{N_m}{c_{{i_2},H}}(1 - {c_{{i_2},H}})} }}{{2{c_{{i_2},H}}(1 - {c_{{i_2},H}})}}}}{{{c_{{i_2},H}}(1 - {c_{{i_2},H}})}} < 0.\\
% &= \frac{{ - \sqrt {M{c_{H,{i_2}}}(1 - {c_{H,{i_2}}})}  - ({c_{H,{i_2}}} - c) (2{c_{H,{i_2}}} - 1)\frac{{\sqrt {M{c_{H,{i_2}}}(1 - {c_{H,{i_2}}})} }}{{2{c_{H,{i_2}}}(1 - {c_{H,{i_2}}})}}}}{{{c_{H,{i_2}}}(1 - {c_{H,{i_2}}})}} < 0\\
\end{split}
\end{align}

The BER generated by the transmitter can be expressed as
\begin{align}
p_e = p_{e,0}+p_{e,1},
\end{align}
where 
\begin{align}
\begin{split}
{p_{e,0}} =&\frac{1}{2}\Bigg[1 - \Phi \left(\frac{{{N_m}{c_{{i_2},L}}}}{\sqrt {{N_m}{c_{{i_2},L}}(1 - {c_{{i_2},L}})} }\right) \\
&\qquad+ \Phi \left(\frac{{{N_m}({c_{{i_2},L}} - c)}}{{\sqrt {{N_m}{c_{{i_2},L}}(1 - {c_{{i_2},L}})} }}\right)\Bigg],
\end{split}
\end{align}
and
\begin{align}
\begin{split}
{p_{e,1}} =& \frac{1}{2}\Bigg[1 - \Phi \left(\frac{{{N_m}(1 - {c_{{i_2},H}})}}{{\sqrt {{N_m}{c_{{i_2},H}}(1 - {c_{{i_2},H}})} }}\right) \\
&\qquad+ \Phi \left(\frac{{{N_m}(c - {c_{{i_2},H}})}}{{\sqrt {{N_m}{c_{{i_2},H}}(1 - {c_{{i_2},H}})} }}\right)\Bigg].
\end{split}
\end{align}
Meanwhile
\begin{align}
\begin{split}
\label{pe0_clb}
\frac{{\partial {p_{e,0}}}}{{\partial {c_{{i_2},L}}}} = \frac{1}{2}\Big( - \frac{{\partial \Phi }}{{\partial A}} \cdot \frac{{\partial A}}{{\partial {c_{{i_2},L}}}} + \frac{{\partial \Phi }}{{\partial B}} \cdot \frac{{\partial B}}{{\partial {c_{{i_2},L}}}}\Big) > 0,
\end{split}
\end{align}
and
\begin{align}
\begin{split}
\label{pe1_chb}
\frac{{\partial {p_{e,1}}}}{{\partial {c_{{i_2},H}}}} = \frac{1}{2}\Big( - \frac{{\partial \Phi }}{{\partial C}} \cdot \frac{{\partial C}}{{\partial {c_{{i_2},H}}}} + \frac{{\partial \Phi }}{{\partial D}} \cdot \frac{{\partial D}}{{\partial {c_{{i_2},H}}}}\Big) < 0.
\end{split}
\end{align}

As shown in (\ref{pe0_clb}) and (\ref{pe1_chb}), $\frac{{\partial {p_{e,0}}}}{{\partial {c_{{i_2},L}}}}>0$ and $\frac{{\partial {p_{e,1}}}}{{\partial {c_{{i_2},H}}}}<0$, meaning that $p_{e,0}$ decreases as $c_{{i_2},L}$ decreases, and $p_{e,1}$ decreases as $c_{{i_2},H}$ increases.
Moreover, the decrease in $c_{{i_2},L}$ leads to an increase in $c_{{i_2},H}$, resulting in both $P_{e,0}$ and $P_{e,1}$ to decrease as $c_{{i_2},L}$ decreases. Consequently, the BER at the transmitter, denoted as $p_e$, decreases with the decrease of $c_{{i_2},L}$. Additionally, a higher number of moving molecules leads to a lower $c_{{i_2},L}$, resulting in a lower BER at the transmitter. 

Among the three movement strategies, continuously moving $i_2$ molecules from the low reservoir to the high reservoir yields the smallest $c_{i_2,L}$ and, consequently, the lowest BER. Cross-moving, where $i_2$ molecules are moved from the low reservoir to the high reservoir while $i_1$ molecules are simultaneously transferred from the high to the low reservoir, achieves a moderately smaller $c_{i_2,L}$ and a correspondingly lower BER. Lastly, moving $i_1$ molecules from the high reservoir to the low reservoir results in the worst BER performance.

\section{Numerical and Simulation Results}
In this section, we first evaluate the transmitter's performance under different movement strategies before the concentration of molecules in the low reservoir reaches equilibrium. We then explore how the number of molecules moved per step influences the transmitter's performance. Finally, we assess the transmitter's performance under various movement strategies after the concentration in the low reservoir has balanced. The simulation parameters are presented in Table 1.
\begin{table}
\normalsize
\caption{SIMULATION PARAMETERS}
\centering
\begin{center}
\renewcommand\arraystretch{1.5} 
\setlength{\tabcolsep}{0.8mm}
\resizebox{0.5\textwidth}{!}{
\begin{tabular}{  c  c  c  p{3cm}}
\hline
Symbol & Explanation & Value  \\ \hline
$k$ & Boltzmann's constant & 1.3807 $\times$ $10^{-23}$ \\ \hline
$T$ & Absolute temperature & 298.15 \\ \hline
$N_m$ & Number of transmitted molecules per bit & $5000$ \\ \hline
$n_L$ & Number of molecules in the $L$ reservoir & $4 \times {10^8}$ \\ \hline
$n_L$ & Number of molecules in the $H$ reservoir & $4 \times {10^8}$ \\ \hline
$n$ & Total number of molecules at the transmitter & $8 \times {10^8}$ \\ \hline
\end{tabular}}
\end{center}
\end{table}
In the simulations, ``Move $i_1$ molecules" refers to transferring $i_1$ molecules from the low reservoir to the high reservoir, while ``Move $i_2$ molecules" refers to transferring $i_2$ molecules from the $L$ reservoir to the $H$ reservoir, and $c$ denotes the initial concentration of $i_2$ molecules in the reservoirs.

\begin{figure}
\centering
\includegraphics[width=0.5\textwidth]{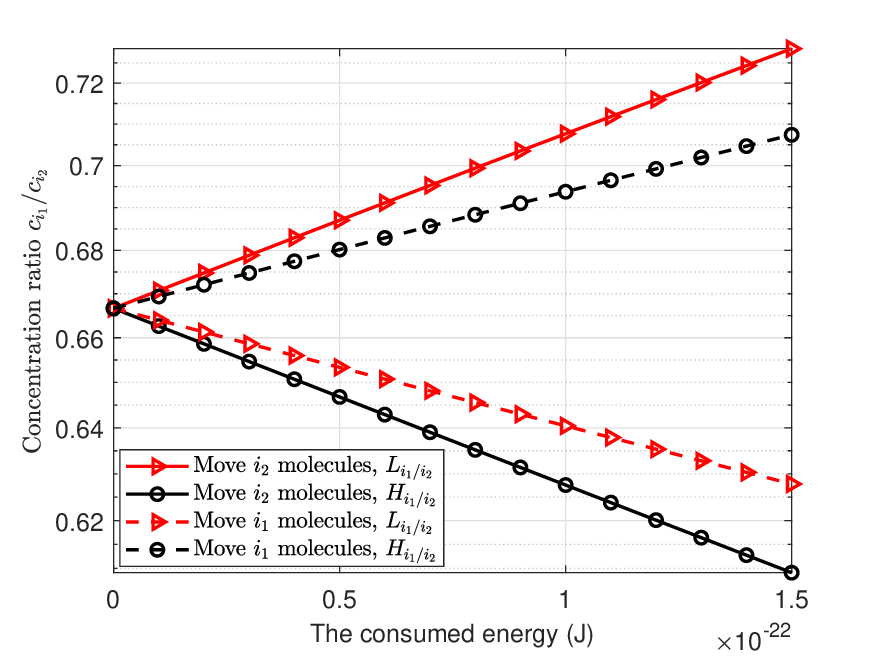}
\caption{The concentration ratios in the reservoirs under different movement strategies.} \label{concentration_ratio_before}
\end{figure}

In Fig. \ref{concentration_ratio_before}, we analyze how the concentration ratio of $i_1$ to $i_2$ molecules in the reservoirs varies with energy consumption under different movement strategies. In the figure, 
$L_{i_1/i_2}$ represents the concentration ratio of $i_1$ to $i_2$ in the $L$ reservoir, and $H_{i_1/i_2}$ represents the concentration ratio of $i_1$ to $i_2$ in the $H$ reservoir.  As energy consumption increases, the movement of $i_1$ molecules causes the concentration ratio of $i_1$ to $i_2$ to decrease in the $L$ reservoir and increase in the $H$ reservoir. In contrast, moving $i_2$ molecules leads to an increase in the concentration ratio of $i_1$ to $i_2$ in the $L$ reservoir and a decrease in the $H$ reservoir. Additionally, since the ``move $i_2$ molecules" strategy involves transferring a larger number of molecules, it results in more pronounced changes in the concentration ratio of $i_1$ to $i_2$ in both reservoirs.

\begin{figure}
\centering
\includegraphics[width=0.5\textwidth]{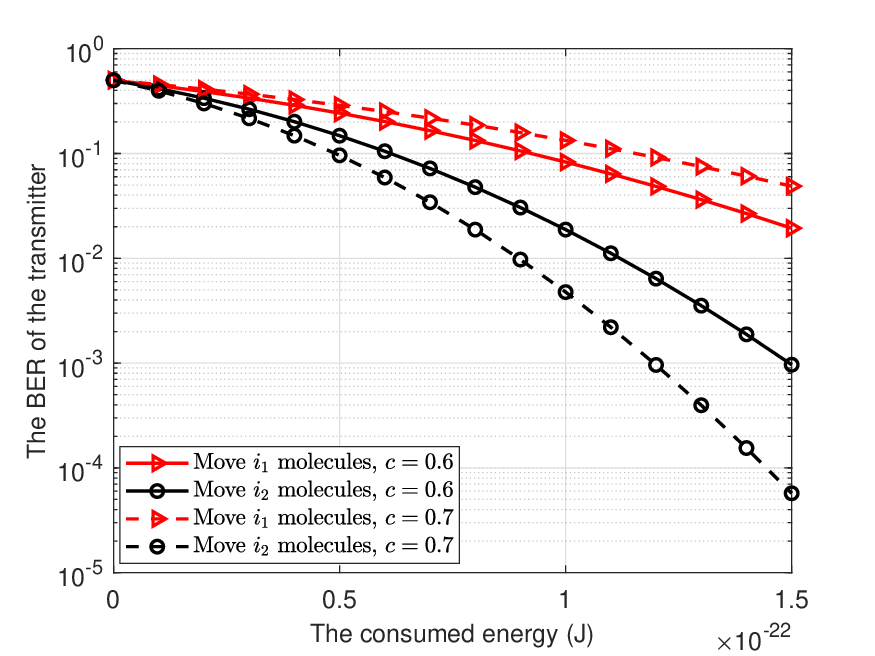}
\caption{The BER of the transmitter under different movement strategies, specifically when the initial concentration of $i_2$ molecules exceeds 0.5.} \label{BER_move_before}
\end{figure}

In Fig. \ref{BER_move_before}, we examine how the BER of the transmitter varies with energy consumption under various movement strategies and initial concentrations of $i_2$ molecules.  The figure illustrates that as energy consumption increases, the BER decreases. This improvement is due to higher energy consumption allowing a larger number of moved molecules, which enhances the gap between different molecular concentrations. 
Moreover, transferring more molecules results in a greater difference in concentration between the reservoirs, leading to a lower BER. As discussed in Section III, when molecules with higher initial concentrations are moved, the number of transferred molecules increases, further reducing the BER of the transmitter.

\begin{figure}
\centering
\includegraphics[width=0.5\textwidth]{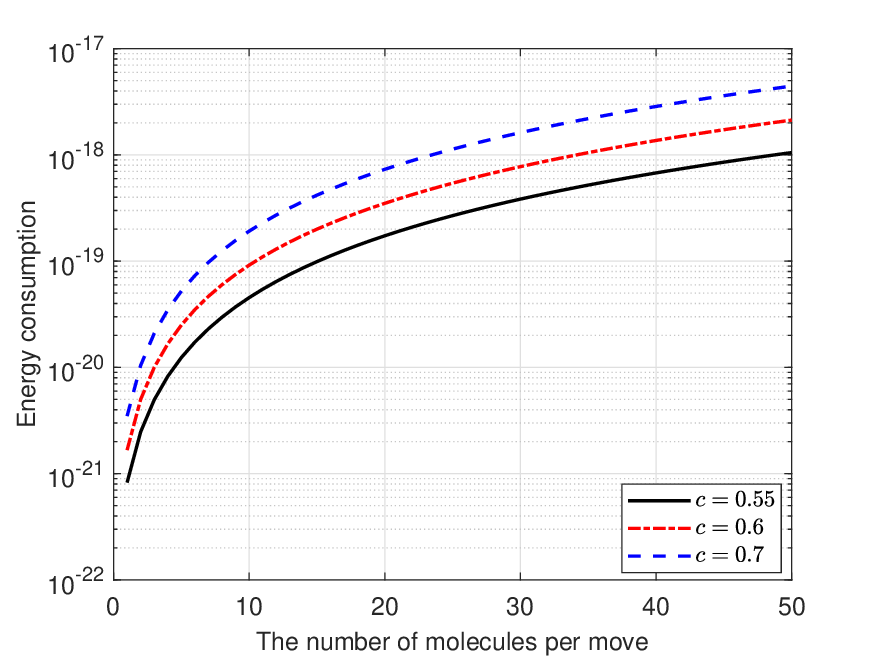}
\caption{Energy consumption varies with the number of molecules moved per step at different initial concentrations of $i_2$ molecules.} \label{Energy_per_move}
\end{figure}

Fig. \ref{Energy_per_move} shows how energy consumption varies with the number of molecules transferred per step at different initial concentrations of $i_2$. The figure illustrates the energy required to move $i_2$ molecules from the low reservoir to the high reservoir until the concentrations of $i_1$ and $i_2$ in the low reservoir reach equilibrium. As the number of molecules moved per step increases, the energy required to transfer the same total amount of $i_2$ molecules also increases. Additionally, at higher initial concentrations of $i_2$ molecules, more molecules must be transferred to achieve equilibrium, which further raises the energy demands.

\begin{figure}
\centering
\includegraphics[width=0.5\textwidth]{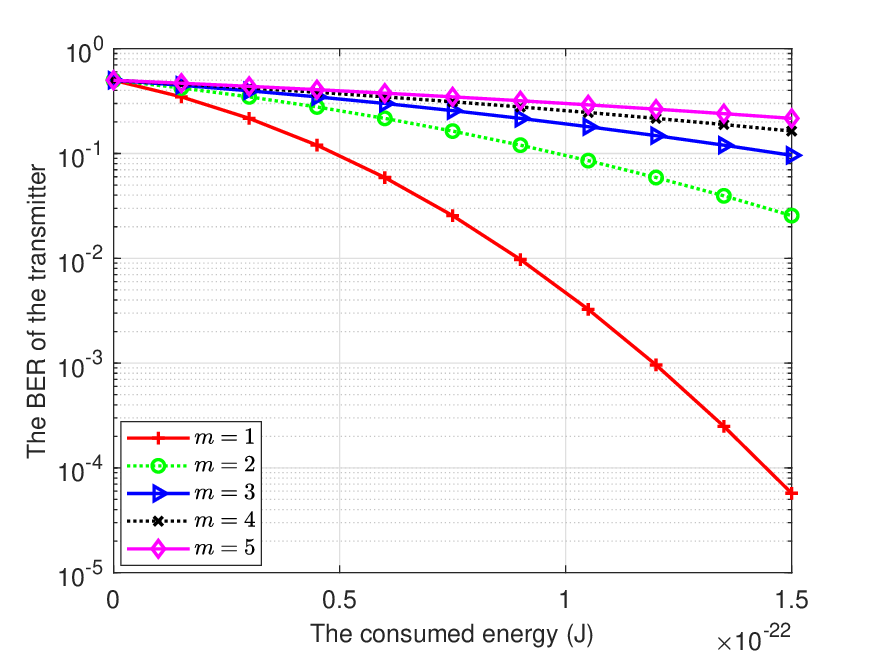}
\caption{The BER of the transmitter varies with energy consumption for different numbers of molecules moved per step.} \label{BER_energy_m}
\end{figure}

Fig. \ref{BER_energy_m} illustrates how the BER of the transmitter varies with energy consumption for different numbers of molecules moved per step. As shown in the figure, the BER decreases with increasing energy consumption because higher energy allows the movement of more molecules, enhancing the concentration ratio and improving BER performance. For smaller values of $m$, as noted in equation (28), less energy is consumed, resulting in even better BER performance. The performance gap widens as $m$ decreases, meaning that moving fewer molecules per step leads to better transmitter performance. Thus, moving a single molecule per step achieves the best transmitter performance.

\begin{figure}
\centering
\includegraphics[width=0.5\textwidth]{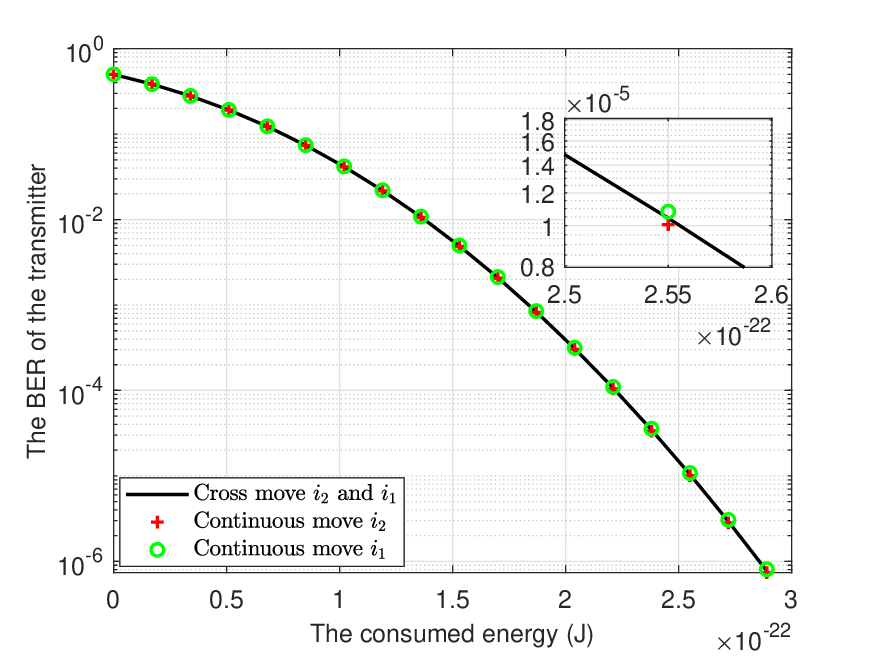}
\caption{The BER of the transmitter varies with the consumed energy under different movement strategies.} \label{BER_energy_strategies}
\end{figure}

Fig. \ref{BER_energy_strategies} analyzes how the BER of the transmitter varies with energy consumption under different molecule movement strategies, after balancing the concentrations of $i_1$ and $i_2$ molecules in the low reservoir. Assuming the initial concentration of $i_2$ molecules is higher than that of $i_1$, and after balancing by transferring $i_2$ molecules from the low reservoir to the high reservoir, we evaluate three movement strategies.
Notably, the strategy of continuous movement of $i_2$ molecules, which involves transferring a larger number of molecules, achieves the best BER performance. In contrast, the strategy of continuous movement of $i_1$ molecules, which moves fewer molecules, results in the worst BER performance. Although the differences in BER performance between the various strategies are relatively small, the results indicate that the continuous movement of $i_2$ molecules is the optimal choice for improving the transmitter’s performance.

\begin{figure}
\centering
\includegraphics[width=0.5\textwidth]{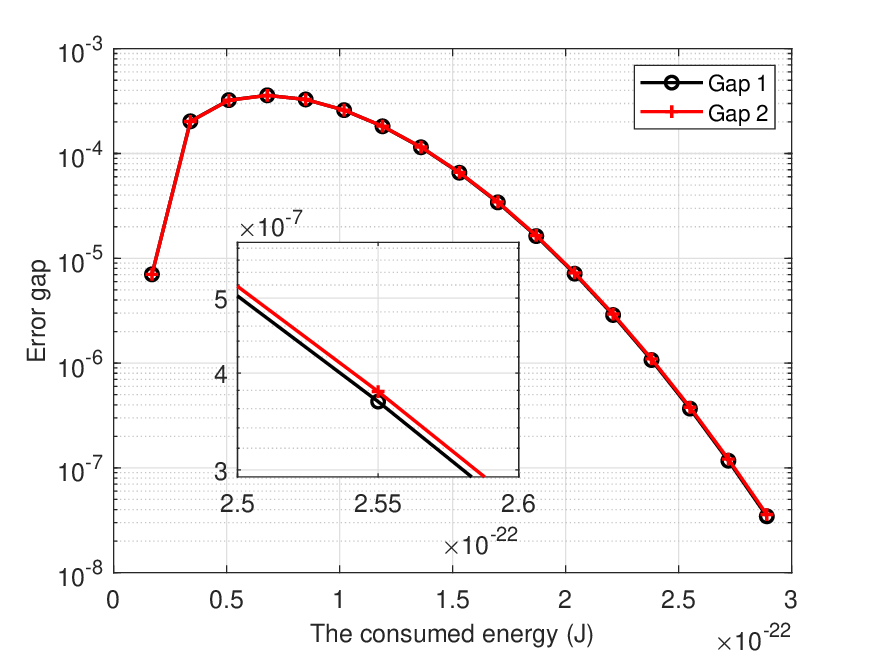}
\caption{The BER gap of the transmitter varies with energy consumption across different movement strategies. Gap 1 represents the BER difference between cross-moving $i_2$ and $i_1$ molecules and moving only $i_2$ molecules. Gap 2 represents the BER difference between moving only $i_1$ molecules and cross-moving $i_2$ and $i_1$ molecules.} \label{BER_energy_strategies_gap}
\end{figure}

Fig. \ref{BER_energy_strategies_gap} shows how the error gap varies with energy consumption under different movement strategies. Initially, the error gap increases as energy consumption rises because fewer $i_2$ molecules are moved at low energy levels, making additional movements more impactful. As energy consumption grows, more molecules are moved, but the increasing energy cost per molecule reduces the gap between strategies. Thus, the performance differences among strategies become minimal. However, consistently moving $i_2$ molecules still yields the best BER performance.

\section{Conclusions}
This paper introduced a method for designing an energy-efficient MC transmitter by moving molecules between reservoirs to create concentration differences for encoding information. Theoretical analysis and simulation results demonstrated that transferring molecules with higher initial concentrations significantly enhanced transmitter performance. Efficiency was further improved by employing a strategy that transferred smaller quantities of molecules per step while achieving a larger cumulative transfer over time. After equilibrium of molecules in the reservoir was reached between the reservoirs, the continued transfer of the initially higher-concentration molecules consistently delivered optimal outcomes. Although the performance differences among the evaluated strategies were marginal, the proposed methods provided valuable insights into optimizing the design of energy-efficient MC systems. A more comprehensive analysis of the end-to-end performance, aimed at evaluating the transmitter's performance, will be presented in future work.
% \end{IEEEbiography}
\bibliographystyle{IEEEtran}
\bibliography{references}
\end{document}